\documentclass[floats,nofootinbib,twocolumn]{revtex4}
\usepackage{amsmath,amssymb,graphicx,latexsym}

\newcommand{\be}{\begin{equation}}
\newcommand{\ee}{\end{equation}}
\newcommand{\bea}{\begin{eqnarray}}
\newcommand{\eea}{\end{eqnarray}}

\newcommand{\bra}[1]{\left|{#1}\right>}
\newcommand{\ket}[1]{\left<{#1}\right|}

\begin{document}
\title{Search for QCD Axions in light of String Theory}

\author{Qiaoli Yang\footnote{Contact author: qiaoliyang@jnu.edu.cn}}
\author{Runchao Huang}

\affiliation{College of Physics and Optoelectronic Engineering, Department of Physics, Jinan University, Guangzhou 510632, China
}


\begin{abstract}
The QCD axion stands as one of the most promising candidates for resolving the strong CP problem. However, the value of the axion's decay constant $f_a$ and, by extension, its mass $m_a$, remain uncertain within the framework of effective field theory, posing a challenge for experimental detection. Fortunately, fields such as cosmology and astrophysics can offer crucial clues about potential mass ranges. Additionally, string theory and the more recent swampland principles might shed some light on the subject. The most straightforward string theory compactifications suggest that $f_a$ is around the GUT scale, leading to a quantum abundance of dark matter. We found that this range can be detected through hydrogen atomic transitions. The recent concept of the dark dimension scenario introduces an alternative possibility. If axions are confined to the four-dimensional Standard Model brane, their decay constant $f_a$ would be on the order of $10^{10}$ GeV. In this scenario, where axions constitute only a minor portion of dark matter, we show that a laser-interferometry setup would be an effective detection method.
\end{abstract}

\date{\today}

\maketitle
\section{Introduction}
The Standard Model QCD vacuum features a parameter $\theta \in [0, 2\pi]$, which leads to a CP-violating topological interaction \cite{tHooft:1976rip}:

\bea
\mathcal{L} = \frac{\theta}{32\pi^2} \text{Tr} G_{\mu\nu} \tilde{G}^{\mu\nu}.
\eea

Although this total derivative term does not contribute to the classical field equations, it has an impact at the quantum level. Experimental limits on the neutron's electric dipole moment \cite{Wurm:2019yfj} suggest $\bar{\theta} \equiv \theta + \text{arg det } m_q < 10^{-10}$, where $m_q$ is the quark mass matrix. The smallness of $ \bar{\theta} $ is known as the strong CP problem. This problem points to a highly fine-tuned relationship between two unrelated quantities, $ \theta $ and arg det $ m_q $. The fine-tuning does not appear to have an anthropic solution, suggesting that a new dynamical mechanism might be required.

A plausible solution to the strong CP problem is the Peccei-Quinn mechanism \cite{Peccei:1977hh,Peccei:1977ur, Weinberg:1977ma, Wilczek:1977pj}. By introducing a dynamical field $ a $ with a shift symmetry $ a \to a + \text{const} $, the $ \theta $ term can be absorbed into the dynamical field, resulting in a new Nambu-Goldstone boson field, the QCD axion. The PQ symmetry is anomalous in the presence of QCD instantons, which explicitly break the shift symmetry and generate a $ \theta $-dependent potential for the axion. The resulting axion field automatically adjusts itself to cancel any possible value of $ \bar{\theta} $, thereby solving the strong CP problem. It's important to note that for the mechanism to work, the axion potential must be dominated by the QCD instanton effect, thus fixing the relationship between the axion mass $ m_a $ and the axion decay constant $ f_a $ as follows \cite{Sikivie:2020zpn}:

\bea
m_a \approx  6 \times 10^{-10} \text{eV} \left(\frac{10^{16} \text{GeV}}{f_a}\right).
\eea

In addition to couplings to gluons, the QCD axion also has couplings to other gauge bosons and fermions, such as \cite{Kim:2008hd}:

\bea
\mathcal{L} \sim \frac{\alpha}{\pi f_a} a F_{\mu\nu} \tilde{F}^{\mu\nu},
\eea

\bea
\mathcal{L} \sim \frac{1}{f_a} \partial_\mu a \bar{f} \gamma^\mu \gamma^5 f,
\eea

where $ F_{\mu\nu} $ are the electromagnetic field tensors, and $ f $ are the fermion fields, respectively. The symbol $\sim$ here indicates that each coupling includes a model-dependent factor of order one.

The uncertainty of the QCD axion primarily stems from the axion decay constant $ f_a $. The earliest model \cite{Weinberg:1977ma} assumed it to be at the electro-weak scale or approximately 250 GeV, but experiments have ruled out this scenario. Subsequent models \cite{Kim:1979if, Shifman:1979if,Zhitnitsky:1980tq, Dine:1981rt} separate the electro-weak scale from the Peccei-Quinn scale, assuming a much higher scale. From the perspective of effective field theory, the uncertainty is quite large. Current laboratory searches and astrophysical constraints typically exclude values of $ f_a $ below $ 10^8 $ GeV \cite{Sikivie:1983ip,CAST:2017uph}. For a more recent review, see Ref.\cite{Ringwald:2024uds}.

Particles with axion-like properties commonly emerge in string theory compactifications as Kaluza-Klein zero modes of antisymmetric tensor fields \cite{Svrcek:2006yi}. For example, the Kaluza-Klein (KK) expansion of the tensor field $ B_{\mu\nu} $ can yield a massless (pseudo) scalar four-dimensional field when both indices are along a cycle. The simplest six-torus compactification has $ \binom{6}{2} = 15 $ distinct two-cycles, resulting in the same number of scalars. Even the most basic realistic string compactifications will produce many more two-cycles. The gauge invariance of the antisymmetric tensor field action results in a zero potential at any order of perturbation theory, thus ensuring that the scalar fields remain massless. However, their Chern-Simons terms can lead to couplings to gauge fields upon KK reduction \cite{Witten:1984dg}, consequently providing a non-perturbative contribution to the potential, thus, these fields acquire a small mass similar to axions. Assuming that one of their potentials is dominated by QCD instantons, the strong CP problem can be resolved. While real-world implementations are much more complex, they can naturally occur within the string theory \cite{Kachru:2003aw}.

String theory and the recent dark dimension scenario offer insights into the axion/axion-like particle decay constant $ f_a $\cite{Svrcek:2006yi,Visinelli:2018utg,Montero:2022prj,Gendler:2024gdo}. Combined with the potential of QCD instantons, the QCD axion mass can be fixed within a relatively narrow range. The remainder of this paper is organized as follows: In Section II, we demonstrate that the hydrogen 1S triplet state is suitable for detecting QCD axions in the general string compactification scenario. In Section III, we show that the interferometry scheme is appropriate for detecting QCD axions in the dark dimension scenario.
\section{General string compactification scenario}
\subsection{Theoretical Considerations for $f_a$ in String Compactifications}
Starting from the ten-dimensional Lagrangian, which is compactified on a six-manifold, the four-dimensional effective Lagrangian for an axion-like particle can be expressed as:
\bea
{\cal L} \supset \frac{f_a^2}{2} (\partial a)^2 - \Lambda^4 U(a),
\eea
where $ U $ is periodic with a period equal to $ 2\pi $, and
\bea\Lambda^4 = \mu^4 \text{e}^{-S}, \eea
$\mu $ represents the ultraviolet (UV) energy scale of the effective theory, and $ S $ originates from the instanton action, typically being larger than 200. The axion decay constant $ f_a $ thus in most string constructions is on the order of \cite{Svrcek:2006yi,Svrcek:2006hf}
\bea f_a \sim \frac{M_{Pl}}{S} \sim 10^{16} \text{GeV} , \eea
where $ M_{Pl} $ is the reduced Planck mass in 4D. Since the $ f_a $ of different axions varies by the area of the compactified two-cycles, they should be in similar order in a general ten-to-four dimension compactification scenario.

When the value of $ f_a $ is much greater than $ 10^{12} $ GeV, one must consider whether the resulting dark matter window is consistent with the isocurvature perturbations of the cosmic microwave background (CMB) radiation \cite{Hertzberg:2008wr,Visinelli:2009zm}. In this case, the dark matter axion field has an initial field value $ a = \theta f_a $ before cosmological inflation. The Gibbons-Hawking temperature during inflation leads to a small standard deviation $ \sigma_\theta f_a $ in the axion field, given by
\bea \sigma_{\theta} f_a \sim \frac{H_I}{2\pi} \,, \eea
where $ H_I $ is the Hubble parameter during inflation. The energy density of the axion particle is
\bea \rho_a \sim \frac{1}{2} m_a \langle a_0^2 \rangle \,, \eea
where the brackets denote the average of the field. After the QCD phase transition, the axion field absorbs energy from the QCD sector while conserving the total energy density, hence
\bea m_a \delta n_a + m_i \delta n_i + 4 \rho_{rad} \left( \frac{\delta T}{T} \right) = 0 \,, \eea
where $ n_a $ is the axion dark matter particle number density, $ n_i $ is the number density of other no-relativistic particles, and $\rho_{rad} $ is the radiation density. The axion field fluctuation results in an observable isocurvature temperature fluctuation. CMB observations require that
\bea \left\langle \left( \frac{\delta T}{T} \right)_{\text{iso}}^2 \right\rangle \lesssim \mathcal{O}(10^{-11}) \,, \eea
which leads to a consistent $ H_I < 10^{12} $ GeV, when $ f_a \sim 10^{16} $ GeV.

Therefore, the dark matter QCD axion window with $ f_a \sim 10^{16} $ GeV (the GUT scale) and $ m_a \sim 10^{-9} $ eV is consistent and particularly intriguing. Indeed, finding QCD axion dark matter in this case would not only explain the strong CP problem but also provide strong support for the concepts of extra dimensions, eternal inflation, and possibly the landscape of string theory more broadly.

\subsection{Detecting GUT Scale Dark Matter Axions with Hydrogen Atoms}
In this section we follow Ref.\cite{Yang:2019xdz} to show that hydrogen atoms can be used to probe the GUT scale cold dark matter axions. Cold dark matter (CDM) particles exhibit a very low velocity dispersion, typically around $10^{-3}$c \cite{Armendariz-Picon:2013jej,Marsh:2015xka}. This characteristic, along with their small mass, leads to an exceptionally high phase space density. This high density can significantly amplify the resonant atomic transitions induced by axions \cite{Sikivie:2014lha}. However, in certain models, such as the KSVZ axions, their interaction with electrons is considerably weaker. Moreover, the typical energy differences in atomic transitions are much greater than the preferred mass of axions.

Therefore, hydrogen atoms are particularly well-suited as targets for GUT dark matter axions. A hydrogen atom consists of one electron and one proton, so a flip in the spin of either the electron or the proton can trigger an atomic transition. This feature allows for a simultaneous investigation of both the axion-electron and axion-proton interactions through a single transition event. Furthermore, the 1S state of the hydrogen atom can be split into a spin singlet and a spin triplet state. By applying a small Zeeman magnetic field, the triplet state can be further divided into three distinct energy levels. The energy difference between the singlet and triplet states is $0.6 \times 10^{-5}$eV, which is suitable for the classical dark matter axion mass window searches. Introducing a micro Tesla Zeeman field can adjust the energy gap between the spin triplet states to be on the order of $10^{-10}$ eV, aligning with the expected energy of GUT QCD axions.

Additionally, hydrogen masers have been extensively utilized, providing a wealth of experience and knowledge regarding the behavior and applications of hydrogen atoms.

The dark matter axion field at the atomic scale can be viewed as a coherent wave with a small velocity dispersion. The average field strength of the axion field is $\langle a_0\rangle \approx {\sqrt{2\rho_{\rm dm}}}/{m_a}$, where $\rho_{\rm dm}$ represents the local dark matter density. The overall dispersion in axion velocities is typically estimated to be $\Delta v \sim 10^{-3}c$. Since the creation mechanism for dark matter axions differs from the thermal production mechanism \cite{Preskill:1982cy,Abbott:1982af,Dine:1982ah,Sikivie:1982qv}, such as that for Weakly Interacting Massive Particles, their velocity dispersion could be even smaller. Some researchers propose that the spread might be as low as $\Delta v \sim 10^{-7}c$ \cite{Armendariz-Picon:2013jej}. The extremely light mass, combined with the small velocity dispersion, results in a very high local dark matter energy spectral density $I_a$:
\bea I_a = \frac{2\rho_{dm}}{m_a\Delta v^2} \,. \eea
This high energy spectral density can greatly facilitate axion searches.

Within atoms, electrons and protons can be treated as non-relativistic, allowing the axion-fermion couplings to be described by the following equation:
\bea H_{i} = \frac{1}{f_a} \sum g_f \left( \frac{\vec p_f \cdot \vec \sigma_f}{m_f} \partial_t a + \vec \sigma_f \cdot \nabla a \right) \,, \eea
where $ f = e, p $ stands for the electron and proton, respectively. Here, $ p_f $ and $ \sigma_f $ represent the momentum and spin operators for the fermions, $ m_f $ is the fermion mass, and $ g_f $ are the couplings to the axion, typically around one in magnitude. Notably in certain models, the axion's coupling to electrons may be significantly weaker, but the coupling to protons remains substantial in order to address the strong CP problem. Additionally, the first term in the equation is proportional to $m_a^2 \bar{r}$, with $ \bar{r} $ being the Bohr radius, and the second term is proportional to $ v m_a $. When $m_a < 10^{-5}$ eV, the first term can be safely dropped \cite{Stadnik:2013raa}. Furthermore, since the wavelength of the incoming axion is much larger than the Bohr radius of the atoms, the equation can be simplified to:
\bea H_i \approx \sum \frac{g_f}{f_a} (\vec \sigma_f \cdot \vec v) m_a \bar{a}_0 \sin(\omega_a t) \,, \eea
where $ \omega_a = m_a (1 + v_a^2 / 2) $ represents the incoming energy of the axion particles. When the energy gap between the atomic states matches the energy of the incoming axions, the induced transition rate $ R $ is
\bea R = \frac{\pi}{f_a^2} \left| \sum g_f \ket{f} (\vec v \cdot \vec \sigma_f) \bra{i} \right|^2 I_a \,. \eea

With a small Zeeman magnetic field applied, the energy splitting between the singlet and the triplet states is given by:
$ \Delta E \approx 2\mu_B B + 5.9 \times 10^{-6} \text{ eV} \,, $
and the energy splitting within the triplet states is
\bea \Delta E'&\approx 3& \times 10^{-6} \text{ eV} [1 - \sqrt{1 + 76.5 \left(\frac{B}{\text{Tesla}}\right)^2} ] \nonumber\\&+&2.6 \times 10^{-5} \text{ eV} \left(\frac{B}{\text{Tesla}}\right) \,. \eea
When the Zeeman field is weak, the splitting between the states $ \bra{1,0} $ to $ \bra{1,1} $ and $ \bra{1,0} $ to $ \bra{1,-1} $ is almost equal, leading to transitions to both states due to the energy spread of axion dark matter. Since at least one of $ g_e $ or $ g_p $ is non-zero, $ \left| \sum g_f \ket{f} (\vec v \cdot \vec \sigma_f) \bra{i} \right|^2 $ is approximately $ v^2 $ for transitions from $ m = 0 $ to $ m = \pm 1 $ states. The event rate $ NR $ of the axion-induced transition is
\bea NR = N \frac{\pi}{f_a^2 m_a} \left(\frac{v}{\Delta v}\right)^2 \rho_{dm} \,. \eea

The primary source of uncertainty is the velocity dispersion $ \Delta v $ of dark matter. For a velocity spread $ \Delta v \sim 10^{-3}c $, an axion mass $ m_a \sim 10^{-9} $ eV, an axion decay constant $ f_a \sim 10^{16} $ GeV, and a quantity of hydrogen atoms $ N $ weighing 1 kg, the event rate $ NR $ is approximately 0.8 per hour. If the velocity spread is as low as $ \Delta v \sim 10^{-7}c $, then for 1 gram of hydrogen atoms, the event rate is 22 per second.

\begin{figure}
\includegraphics[width=0.4\textwidth]{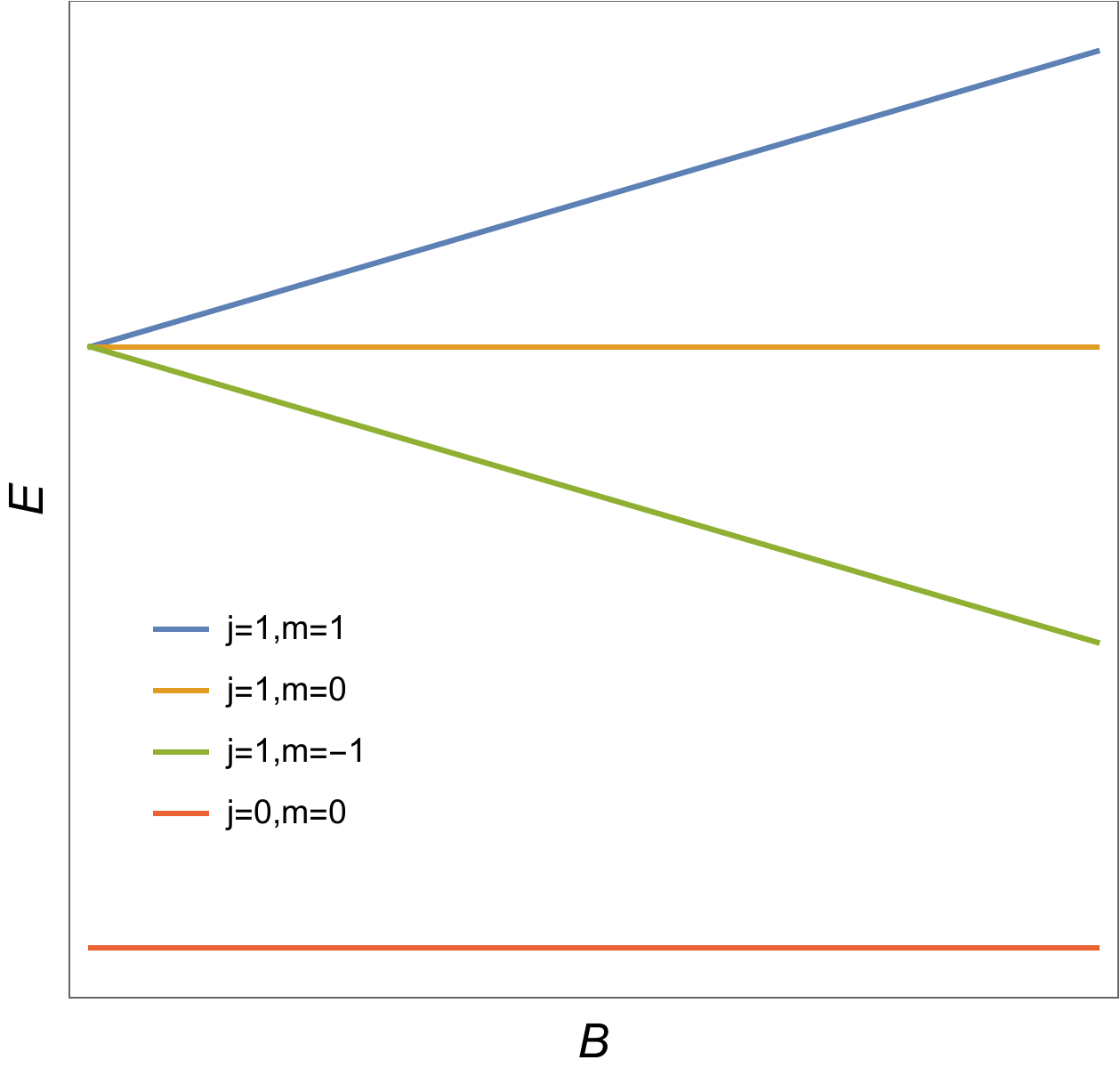}
\caption{The Zeeman splitting of the ground state of hydrogen atom.}
\end{figure}

\begin{figure}
\includegraphics[width=0.4\textwidth]{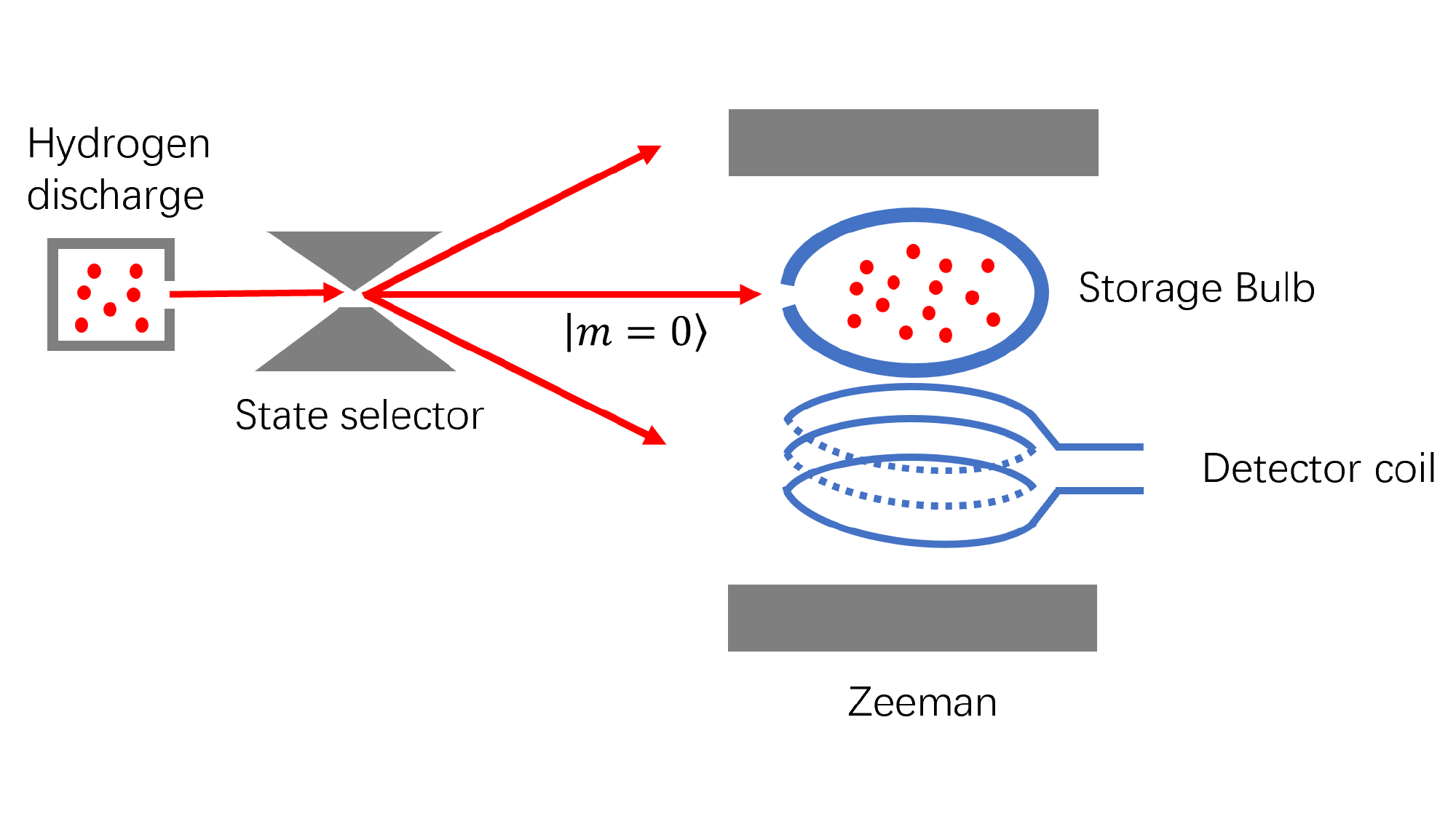}
\caption{A possible experimental setup that could be similar to existing masers.}
\end{figure}

\begin{figure}
\includegraphics[width=0.5\textwidth]{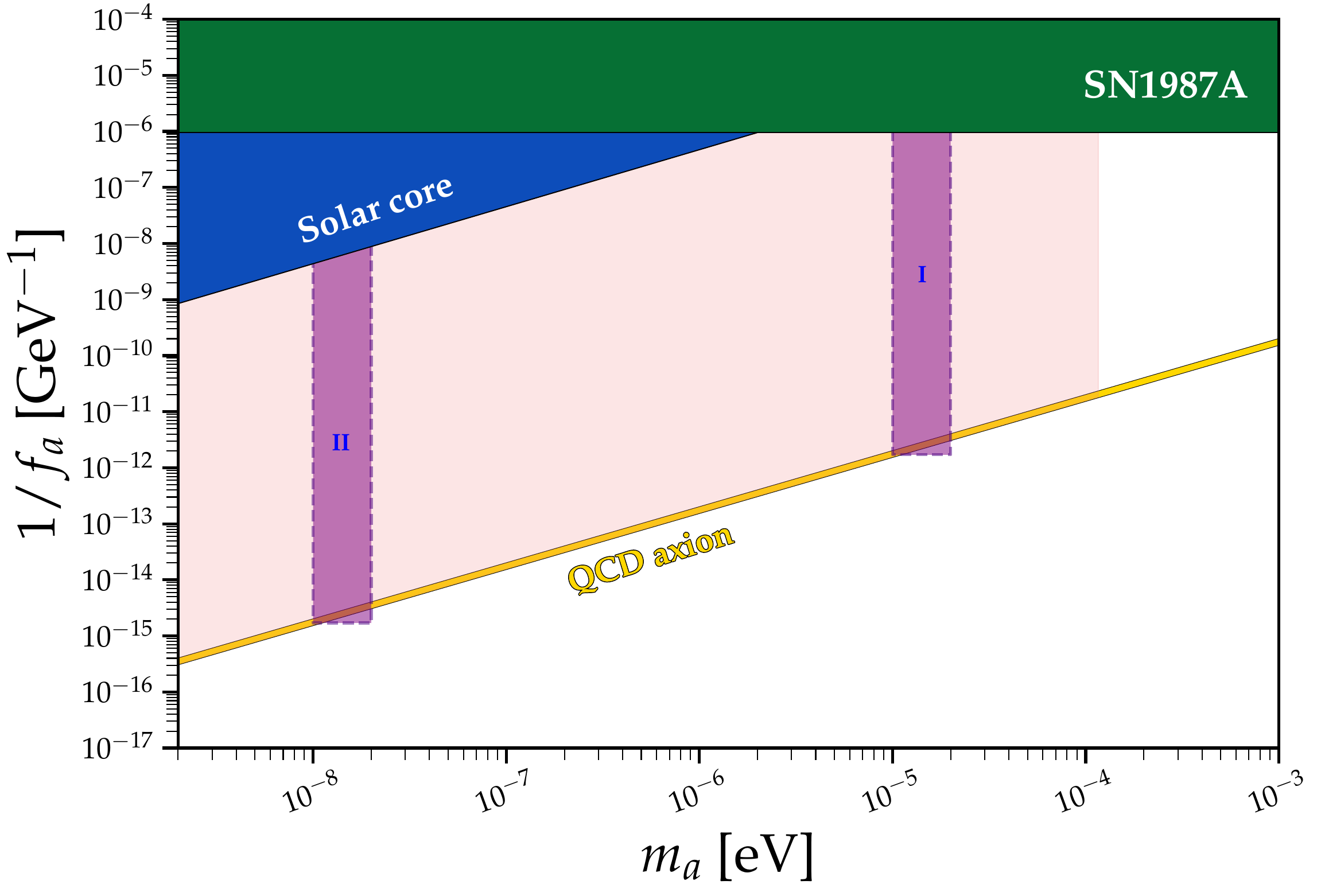}
\caption{The mass range that can be covered is determined by the available Zeeman field strength. To cover Region I (\( m_a \in [10^{-5} \, \text{eV}, \, 2 \times 10^{-5} \, \text{eV}] \)), a field strength of 0.15 Tesla is needed. Assuming 10\% of this region is covered in a one-year data-taking period using 1 mole of H atoms, the sensitivity is \( f_a = 5.8 \times 10^{11} \, \text{GeV} \). For Region II (\( m_a \in [10^{-8} \, \text{eV}, \, 2 \times 10^{-8} \, \text{eV}] \)), a field strength of 0.002 Tesla is required. Assuming 10\% of this region is covered using \( 10^3 \) moles of H over a one-year data-taking period, the sensitivity is \( f_a = 5.8 \times 10^{14} \, \text{GeV} \). The yellow band defines the QCD axion parameter space. The solid-colored regions, SN1987A and Solar Core, have been excluded by experiments and observations. The light coral region can be probed using this hydrogen splitting method if a 1 Tesla Zeeman field and an adequate amount of hydrogen atoms are deployed.}
\end{figure}

Since the exact value of the axion mass is not known, the experiment must scan the possible mass range, which can be achieved by adjusting the Zeeman magnetic field $ B $. The scanning rate depends on the mass range covered. For example, to cover a mass range of $ \Delta f = 10\% \times 10^{-9} \text{ eV} / 2\pi \sim $ 0.1 MHz in one year, the scanning rate would be $ R_s = \Delta f / \text{one year} = 0.00317 $ Hz/s.

Because the inherent bandwidth of the dark matter axion is $ \Delta f_{dm} = m_a \Delta v^2 / 2 $, so the effective integration time during scanning at each frequency is $ t_{int} = \Delta f_{dm} / R_s $. Assuming a 3-sigma detection criterion, $ NRt_{int} \geq 3 $, the sensitivity is then:
\bea f_a \leq 0.9 \times 10^{15} \text{ GeV} \sqrt{\frac{0.1 \text{ MHz}}{\Delta f}} \sqrt{\frac{N}{10^3 \text{ mole}}} \,. \eea

\section{dark dimension scenario}
\subsection{Theoretical Considerations of $f_a$ in The Dark Dimension Scenario}
Effective quantum field theory currently forms the core of fundamental physics, yet they might include an overabundance of degrees of freedom. Fortunately, the fine-tunings observed in the parameters of these models could hint a path to a more fundamental theory. Additionally, just as the spin-orbit interaction arises from the Dirac equation, it is reasonable to anticipate that very high-energy physics (UV) could have consequences at lower energies (IR). As a result, some effective field theories might be inconsistent with the UV theory of quantum gravity. These considerations have led to the development of the Swampland principles \cite{Vafa:2005ui} .

The Swampland principles, when paired with the observed smallness of dark energy, suggest the dark dimension scenario \cite{Montero:2022prj}. This scenario predicts the existence of an extra dimension within the range of 1 to 10 microns, which could present fascinating phenomenology. The scale of the additional dimension, denoted as $ R_d $, is related to dark energy, $ \Lambda $, by the formula $ R_d \sim \Lambda^{-1/4} $. Given that the mass of the Kaluza-Klein (KK) states in such a large extra dimension would be on the order of meV, the Standard Model particles must be confined to a 4D brane to be consistent with experimental observations.

Axions, on the other hand, can either reside on the 4D Standard Model brane or they can traverse both the 4D brane and the dark dimension, which makes up the 5D bulk. In the latter case, the axion decay constant would be similar to that observed in typical string compactifications. A unique phenomenological feature is the presence of an additional KK tower of axions, with their masses determined by the radius of the dark dimension. The existence of this KK tower of axions does not contradict current observational constraints. However, this scenario seems less likely for QCD axions, as they are closely tied to processes within the Standard Model.

When the QCD axion is confined to the Standard Model brane, considering a generalization of the Weak Gravity Conjecture and constraints from the cooling of neutron stars, the decay constant of the QCD axion is estimated to be approximately $ f_a \sim 10^9 - 10^{10} $ GeV \cite{Gendler:2024gdo, Vafa:2024fpx}. The cold dark matter axions that result from the most natural misalignment mechanism would account for only about one percent of the total dark matter in this scenario. Therefore, searches for axions that do not depend on dark matter are more suitable in this context.

\subsection{Detecting Dark Dimension Scenario Axions by Interferometry}
In this section we follow Ref.\cite{Tam:2011kw} to show that dark dimension scenario axions can be probed by interferometry. Axion searches that do not depend on dark matter often concentrate on astrophysical sources. Experiments such as the CERN Axion Solar Telescope (CAST) are examples of helioscopes \cite{Sikivie:1983ip}, designed to detect axions emitted by the Sun by converting them into X-rays within a strong magnetic field. Pure laboratory based experiments are possible. For example, the interaction between photons and axions can also lead to birefringence and dichroism \cite{Ahlers:2008jt}, causing the polarization of light to rotate and become elliptical in the presence of a magnetic field. Experiments looking for these effects include polarimetry studies such as BFRT and PVLAS, among others. Another type of experiment that leverages this mixing is photon-regeneration \cite{VanBibber:1987rq}, where a part of a laser beam goes through a magnetic field, converting to axions. Because axions have a weak interaction with normal matter, they can pass through a wall without hindrance. On the other side of the wall, there is an arrangement of magnets that can cause some of the axions to convert back into photons. The main advantage of these experiments is better control and understanding of the physical processes, as they are not dependent on extraterrestrial sources. However, the signal is typically weak due to the large background (polarimetry) or due to the requirement for two conversion stages (photon-regeneration).

To address these issues, a new experimental method based on interferometry has been proposed \cite{Tam:2011kw}. In this setup, a laser beam is split into two, with one serving as a reference beam and the other passed through a magnetic field to trigger conversion into axions. The beam is then recombined with the reference beam. If axion conversion has occurred, the beam from the conversion area will have a reduced amplitude and a phase shift relative to the reference beam. This results in a change in the combined intensity, which can be measured. Since only one conversion stage is needed, the signal is improved.

To ensure that the signal is not drowned out by background noise, the two beams could be arranged to travel paths of different optical lengths, causing them to go out of phase by $\pi$ when the magnetic field is off. Thus the two beams would interfere destructively at the detector if no photons are converted into axions. The disadvantage is that at the dark fringe, the signal is reduced to a second-order effect. Adding a modulation of the laser's amplitude can overcome this issue. However, using a coherent laser light source still introduces shot noise. For an incoming laser beam containing $ N $ photons, the Heisenberg uncertainty principle causes a fluctuation of $ \sqrt{N} $ in the photon count. This limits its ability to detect axions. On a positive note, the interferometry setup allows for a straightforward implementation of light squeezing, which can reduce shot noise significantly. Moreover, by adding optical delay lines or Fabry-Perot cavities, the scheme can additionally amplify the signal by a factor of $ n $, where $ n $ is the number of repetitions of a laser beam in the optical delay line or FP cavity. The increased sensitivity to axion-photon coupling is a factor of $ n^{1/2} $. In contrast, using an optical delay line or Fabry-Perot cavity in photon-regeneration experiments results in a less pronounced enhancement. Intriguingly, with these improvements, the interferometry scheme with a modest setup can probe the dark dimension axion parameter space.

\begin{figure}
\includegraphics[width=0.4\textwidth]{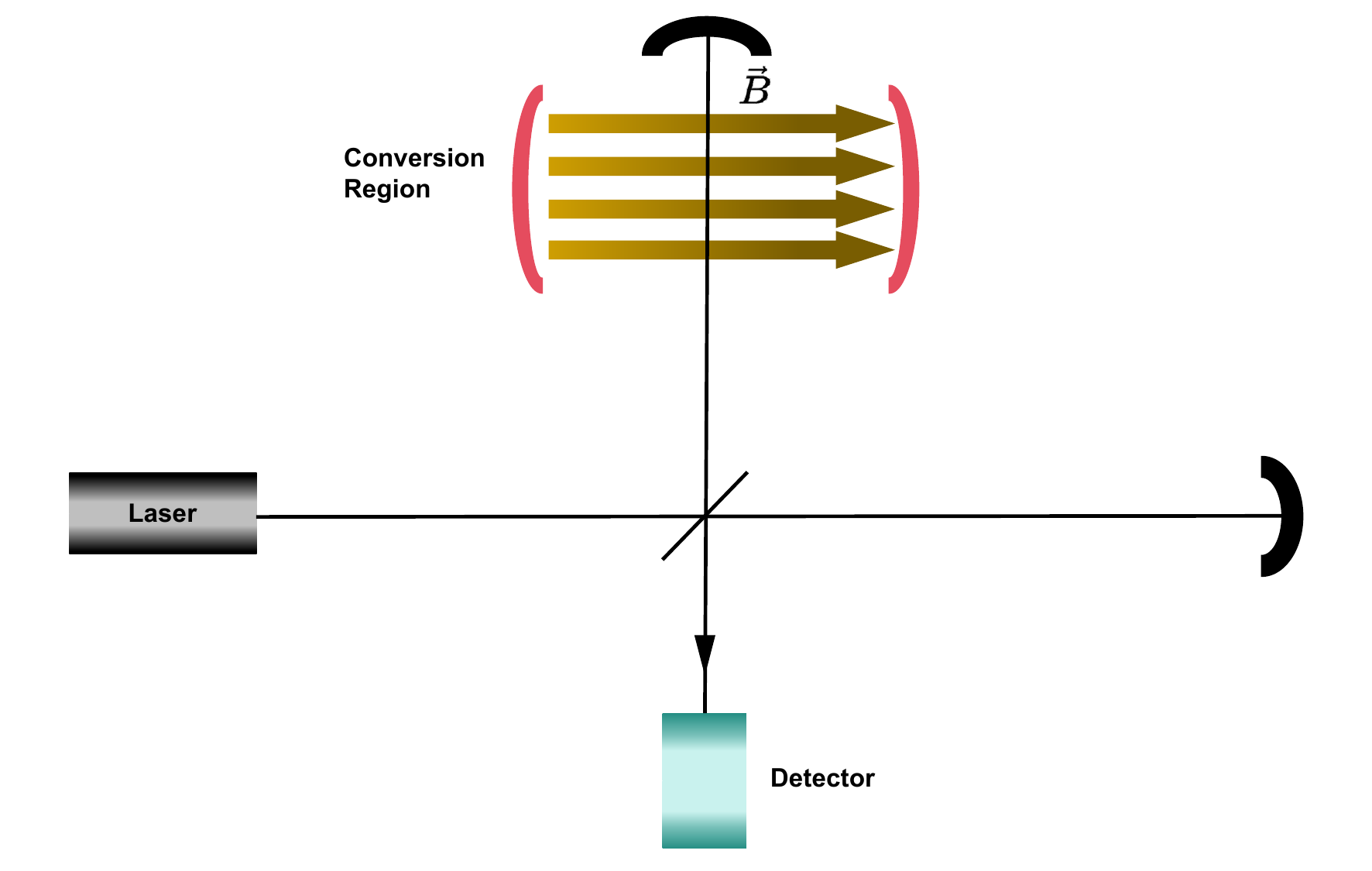}
\caption{Detecting dark dimension axions via interferometry.}
\end{figure}

\begin{figure}
\includegraphics[width=0.5\textwidth]{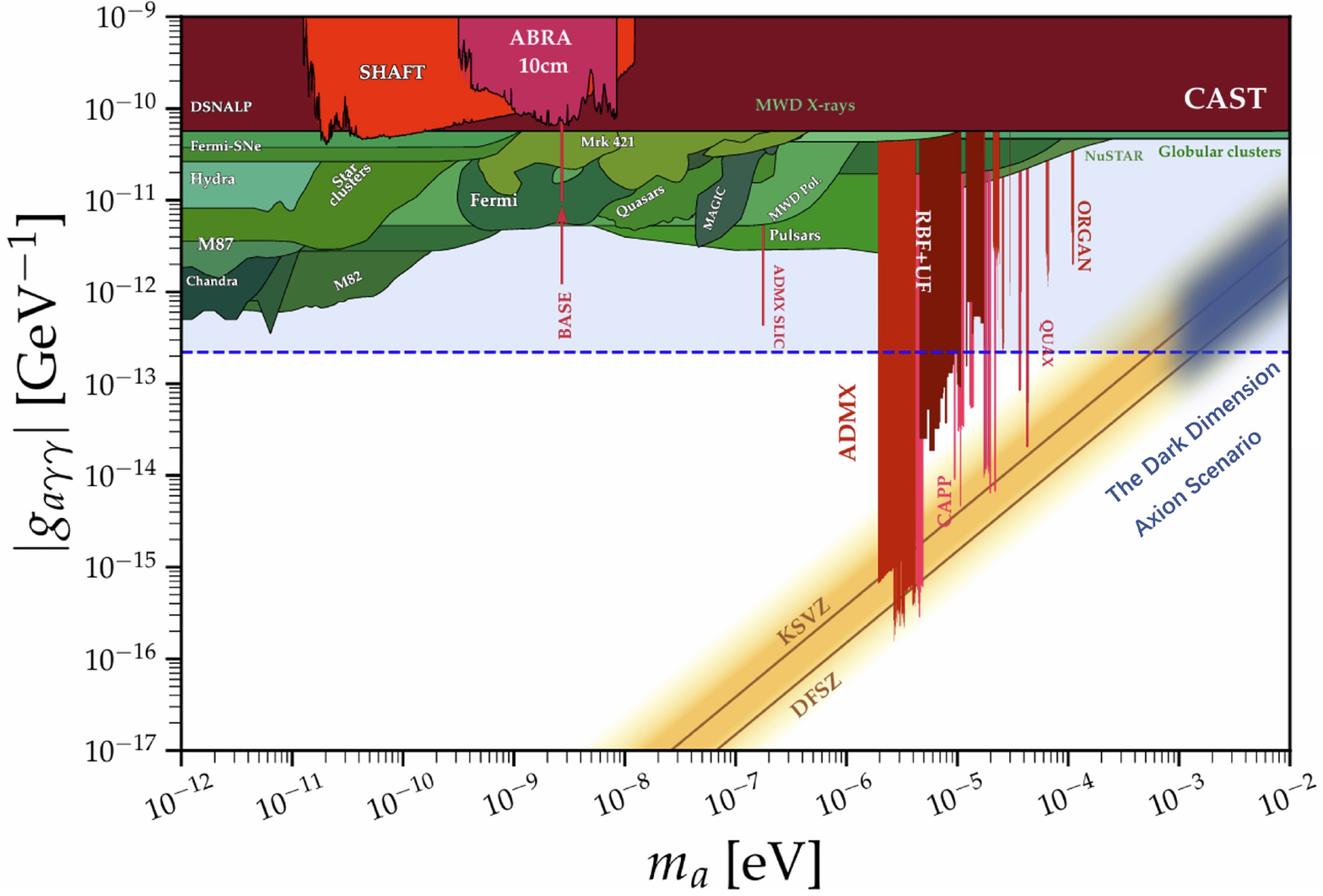}
\caption{Assuming \( n \sim 10^6 \), \( B \sim 40 \) T, \( L \sim 30 \) m with a 10 W (\( \lambda = 1 \, \mu \)m) laser, and squeezed light achieving approximately \({\cal O}(10)\) dB shot noise reduction, the laser experiment can, after 10 days of operation, probe axions with \( g_{a\gamma\gamma} \ge 2.3 \times 10^{-13} \, \text{GeV}^{-1} \) at a \( 5\sigma \) significance level.}
\end{figure}

When photons travel through an area with a magnetic field, if the photon's polarization is aligned with the magnetic field, they transform into axions. The conversion rate is \cite{Sikivie:1983ip,Sikivie:1985yu}
\bea
\eta_{\gamma \rightarrow a} = \frac{1}{4 v_a}(g_{a\gamma\gamma} B L)^2\left(\frac{2}{qL}\sin{\frac{qL}{2}}\right)^2,
\eea
where $ v_a $ is the velocity of the axion, $ B $ is the strength of the magnetic field, $L $ is the length of the conversion region, and $ q $ is the transferred momentum. Given that the axion mass is significantly less than the energy of laser photons, the velocity of the converted axion is nearly the speed of light, and the transferred momentum $ q \approx {m_a^2}/{2\omega_{\gamma}} $. For $ L $ at approximately 10 meters and $ m_a < 10^{-3} $ eV, we have $ qL \ll 1 $. The conversion rate can be approximated as:
\bea
\eta_{\gamma \rightarrow a} \approx \frac{1}{4}(g_{a\gamma\gamma}B L)^2,
\eea
which indicates that the rate increases as the length of the conversion region increases. After conversion, the amplitude $ A $ of the laser field decreases as follows:
\bea
\delta A_{\gamma \rightarrow a} \approx \frac{g_{a\gamma\gamma}^2 B^2 L^2 A}{8}.
\eea
This reduction is valid when $ m_a \ll \sqrt{{2\pi\omega_{\gamma}}/{L}}\equiv m_0 $. If the axion mass is close to or heavier than $m_0$, the axion-photon conversion effect decreases rapidly. To restore sensitivity, the conversion region can be filled with appropriate media such as an inert gas to effectively reduce the photon wavelength, after which the conversion rate can be resumed.

Additionally, these photons also experience a phase shift \cite{Raffelt:1987im}. When $ g_{a\gamma\gamma}^2B^2 \ll m_a^2 $, the phase shift relative to photons that have traveled the same distance $L$ is:
\bea \delta \theta \approx \frac{g_{a\gamma\gamma}^2 B^2 \omega_{\gamma}^2}{m_a^4}\left(\frac{m_a^2L}{2\omega_{\gamma}} - \sin\left(\frac{m_a^2L}{2\omega_{\gamma}}\right)\right), \eea
which has a negligible effect compared with the amplitude change $\delta A/A$. Furthermore, when amplitude modulation technique is used, the detector receives the first-order signal due to the amplitude change but only the second-order signal due to the phase shift (see Eq.(\ref{power})). Notably even without conversion to axions, the QED vacuum in the presence of a magnetic field is inherently birefringent due to loop corrections (the Heisenberg-Euler term) \cite{Heisenberg:1936nmg, Schwinger:1951nm, Dobrich:2009kd}. For $ B \sim 10 $ T, $ L \sim 10 $ m, $ \omega_{\gamma} \sim $ eV, the QED-induced phase shift is also negligible.

To ensure the signal is not drowned out by background noise, the two beams are arranged to be interfere destructively at the detector. The disadvantage is the signal is also reduced to a second-order effect, but one could add an amplitude modulation to the laser to overcome this issue. Assuming the amplitude modulation is at a frequency $ \omega_m $, the laser beam can be written as:
\bea \vec E _{in}= \vec E_{0} (1+\beta \sin \omega_m t) e^{i\omega t}, \eea
where $ \beta $ is a constant and $ \omega $ is the laser photon frequency. The modulation frequency can be chosen to maximize the signal as $ \omega_m \approx n\pi c/2\Delta L $ where $ n $ is an odd integer, and $ \Delta L $ is the optical length difference between the two arms of the interferometer.

The power $ P $ that reaches the detector is
\bea P&=&P_{in}\big( \frac{(\delta A/A)^2+\delta \theta ^2}{4} + \frac{\beta^2(4-4\frac{\delta A}{A}+\frac{\delta A^2}{A^2}+\delta\theta ^2)}{2} \nonumber \\ &+& \beta \left(2\frac{\delta A}{A}- \frac{\delta A^2}{A^2}+\frac{\delta\theta^2}{2}\right)\cos \left[\omega_m \left(t + \frac{2 L}{c}\right)\right] \nonumber\\&+& \frac{\beta^2(4-4 \frac{\delta A}{A}+ \frac{\delta A^2}{A^2}+\delta \theta ^2)}{2}\cos \left[2\omega_m \left( t + \frac{2L}{c}\right)\right]\big),\nonumber\\
\label{power}\eea
where $ L=(L_1+L_2)/2 $ is the average of the two arms of the interferometer. We see that the power has direct current components and alternating current components. One can extract the alternating part experimentally via a mixer. The time-averaged output power with frequency $ \omega_m $ is
\bea P_{out} = \frac{1}{T} \int_T 2 P_{in}\beta \mathcal{G} \left(\frac{\delta A}{A}\right) \cos^2 \left(\omega_m t\right) = \frac{P_{in} \beta\mathcal{G}\delta A}{A} \eea
where $ \mathcal{G} $ is the gain of the detector and $ T $ is the integration time. The sensitivity is restored to be proportional to $ \delta A $.

The utilization of interferometers, however, comes with the presence of shot noise, which reduces the sensitivity to detecting axions. For a laser beam comprising $ N $ incoming photons, the shot noise is $ \sqrt{N} $. Consequently, the signal-to-noise ratio is reduced to $ (g_{a\gamma\gamma}BL)^2 N/\sqrt{N} $. This converts to a sensitivity of axion coupling $ g_{a\gamma\gamma} < (BL)^{-1}N^{-1/4} $. An implementation of squeezed light can assist in mitigating shot noises. When the interferometer operates near a dark fringe, the shot noise can be theoretically reduced indefinitely. Implementations of squeezed light are demonstrated with a $ 15 $ dB shot noise reduction \cite{Vahlbruch:2016qwp}. Additionally, by adding optical delay lines or Fabry-Perot (FP) cavities, the signal can be further enhanced by a factor of \( n \), where \( n \) is the number of times the laser beam is repeated inside the device. Equivalently, this results in an \( n^{0.5} \) improvement in our ability to constrain \( g_{a\gamma\gamma} \).

By employing \( n \sim 10^6 \), \( B \sim 40 \) T, \( L \sim 30 \) m with a 10 W (\( \lambda = 1 \, \mu \)m) laser, and squeezed light achieving approximately \({\cal O}(10)\) dB shot noise reduction, the experiment can, after 10 days of operation, probe axions with \( g_{a\gamma\gamma} \ge 2.3 \times 10^{-13} \, \text{GeV}^{-1} \) at a \( 5\sigma \) significance level. 

\section{Conclusions}
While the standard model of particle physics has achieved great success, the pursuit of a deeper understanding of our universe could benefit from both theoretical and experimental searches for axions. The weak interactions and low mass of axions are a result of very high ultraviolet (UV) energy physics, making axion searches a way to explore the UV energy region within current technology. We demonstrate that GUT-scale QCD axion dark matter can be investigated using hydrogen atoms, and axions in the dark dimension scenario can be explored through a specialized interferometer experiment.

\section*{Acknowledgments}
This work was supported in part by the NSFC under grant no.12150010.

\bibliography{refs}

\end{document}